\documentclass[conference]{IEEEtran}

\usepackage{gensymb}
\usepackage{fancyhdr}
\usepackage{amsmath}
\usepackage{amssymb}
\usepackage{graphicx}
\usepackage{booktabs}
\usepackage{amsfonts}
\usepackage{multirow}
\usepackage{algorithm}
\usepackage{algorithmic}
\usepackage{mathrsfs}
\usepackage{bm}
\usepackage{exscale}
\usepackage{relsize}
\usepackage{setspace}
\usepackage{color}
\usepackage{comment}
\usepackage{ulem}
\usepackage{cite}
\usepackage{geometry}
\newtheorem{remark}{Remark}

\DeclareMathOperator*{\argmax}{arg\,max}

\ifCLASSINFOpdf
\else
\fi

\begin{document}
	
	\title{Triple-Structured Compressive Sensing-based Channel Estimation for RIS-aided MU-MIMO Systems}

	\author{
	Xu~Shi\textsuperscript{1},~Jintao~Wang\textsuperscript{1,2},Guozhi~Chen\textsuperscript{1},~Jian~Song\textsuperscript{1,2}\\
	\IEEEauthorblockA{
	\textsuperscript{1}Beijing National Research Center for Information Science and Technology (BNRist),\\
	Dept. of Electronic Engineering, Tsinghua University, Beijing, China\\
	\textsuperscript{2}Research Institute of Tsinghua University in Shenzhen, Shenzhen, China\\
	\{shi-x19@mails., wangjintao@, chen-gz19@mails., jsong@\}tsinghua.edu.cn}}
	\maketitle
	
	\IEEEoverridecommandlockouts
	\begin{abstract}
		Reconfigurable intelligent surface (RIS) has been recognized as a potential technology for 5G beyond and attracted tremendous research attention. However, channel estimation in RIS-aided system is still a critical challenge due to the excessive amount of parameters in cascaded channel. The existing compressive sensing (CS)-based RIS estimation schemes only adopt incomplete sparsity, which induces redundant pilot consumption. In this paper, we exploit the specific triple-structured sparsity of the cascaded channel, i.e., the common column sparsity, structured row sparsity after offset compensation and the common offsets among all users. Then a novel multi-user joint estimation algorithm is proposed. Simulation results show that our approach can significantly reduce pilot overhead in both ULA and UPA scenarios. 
	\end{abstract}
	
	\begin{IEEEkeywords}
		Reconfigurable intelligent surface (RIS), cascaded channel estimation, structured compressive sensing
	\end{IEEEkeywords}
	
	\IEEEpeerreviewmaketitle
	
	\section{Introduction}
	
	Recently, reconfigurable intelligent surface (RIS) has been recognized as a potential technology for 5G \& beyond and attracted tremendous research attentions \cite{RIS_1,RIS_2}. As a new electromagnetic (EM) material equipped with integrated electronic circuits, RIS elements can independently reflect incident signals by controlling phases to form a strong energy focusing without additional power amplifiers. Due to the passive reflection paradigm, RIS displays great advantages in the low hardware cost and energy consumption compared with regular Amplify-and-Forward (AF) relays \cite{RIS_benefit}. More importantly, additional non-line-of-sight (NLoS) paths can be complemented via the deployment of RIS, especially when the line-of-sight (LoS) channel is obstructed by buildings or trees. In another word, RIS provides an opportunity to change channels actively and helps to enhance the wireless cellular network coverage.
	
	Unfortunately, it is a serious challenge to acquire the full channel state information (CSI) in RIS-assisted system. Since no radio frequency (RF) chains are equipped at the passive RIS, CSI can only be estimated through active antennas at base station (BS) and user equipment (UE) side \cite{RIS_1,RIS_2,RIS_benefit}. Besides, compared with conventional direct channel from UE to BS, the channel in RIS-assisted system contains a quite larger dimension, as a compound of the channel between RIS and BS, and channels between UEs and RIS. 
	
	To the best of our knowledge, cascaded channel estimation was first studied in \cite{Onoff_est}, where RIS turns on only one element successively in every timeslot for estimation. Another decomposition-aided RIS channel estimation approach was then proposed in \cite{zhouzhengyi}. However in above schemes, the pilot overhead is proportional to the size of RIS elements and turns unacceptable when massive RIS elements are equipped. Therefore, sparsity-assisted compressive sensing methods \cite{OMP_conventional,weixiuhong,chen2019channel,matrixcalibration} were further studied to pursuit lower overhead consumption. Furthermore, some sparse structures were partially analyzed in \cite{chen2019channel,weixiuhong} including the common RIS-BS sparsity after subspace projection \cite{chen2019channel}, and a courageous assumption that all users share partially common paths \cite{weixiuhong}. But unfortunately, to our best knowledge, no estimation approach has completely analyzed the specific structured sparsity in RIS cascaded channel, which is greatly helpful to reduce pilot overhead.
	
	In this paper, by exploiting the triple-structured sparsity of the cascaded beamspace channel, we propose a novel  compressive sensing (CS)-based multi-user joint estimation algorithm. Regardless of the column sparsity which has been widely used in previous studies \cite{chen2019channel,weixiuhong}, two another structures are first analyzed in this paper, i.e., structured row sparsity after offset compensation and the common offsets among all users. Furthermore, we extend the algorithm from uniform linear array (ULA) configuration to uniform planar array (UPA). Simulation results show that our proposed algorithm reduces pilot overhead significantly and outperforms several previous partial-structured CS-based approaches.

	\section{System Model}
	In this paper we consider a RIS-aided MU-MIMO uplink narrow-band system, as shown in Fig.1, where $K$ single-antenna users transmit signals to one base station (BS) through RIS. The BS and RIS are equipped with $N_\text{BS}$ antennas and $N_{I}$ reflective elements respectively, which are inhere assumed as ULA \footnote{Actually UPA model is more suitable to reflective elements. The extension to UPA will be shown in Section V.} for convenient analysis. $\mathbf{G}\in \mathbb{C}^{N_\text{BS}\times N_\text{I}}$ denotes the channel matrix from $N_\text{I}$ reflective elements to the BS, and $\mathbf{h}_\text{k}$ denotes the channel from the $k$-th user to the RIS ($k=1,\dots,K$). Then the channel matrices $\mathbf{G}$ and $\mathbf{h}_\text{k}$ can be represented via Saleh-Valenzuela channel model as follows:
	\begin{equation}
	\mathbf{G}=\sum_{l_1=1}^{L_1}\alpha_{l_1}^{\text{G}}\mathbf{a}_{N_\text{BS}}\left(\sin(\theta_{l_1}^{\text{G}_r})\right) \mathbf{a}_{N_\text{I}}\left(\sin(\theta_{l_1}^{\text{G}_t})\right)^H,
	\label{RIS_BS_channel}
	\end{equation}
	\begin{equation}
	\mathbf{h}_\text{k}=\sum_{l_2=1}^{L_2^{(k)}}\alpha_{l_2,k} \mathbf{a}_{N_\text{I}}\left(\sin(\theta_{l_2})\right),k=1,\dots,K,
	\label{UE_RIS_channel}
	\end{equation}
	where $\alpha_{l_1}^\text{G}$ and $\alpha_{l_2,k}$ denote  gains for the $l_1$-th path of RIS-BS channel and the $l_2$-th path of $k$-th UE-RIS channel. $L_1$ and $L_2^{(k)}$ are the total path numbers for channel matrices $\mathbf{G}$ and $\mathbf{h}_k$. The angles of arrival (AoAs) and departure (AoDs) for RIS-BS channel are marked as $\theta_{l_1}^{\text{G}_r}$ and $\theta_{l_1}^{\text{G}_t}, \ l_1=1,\dots,L_\text{1}$. Similarly $\theta_{l_2}$ represents AoA of the $l_2$-th  UE-RIS path for the $k$-th user. $\mathbf{a}_N(\cdot)$ denotes the ULA array steering vector and can be formulated as:
	\begin{equation}
	\mathbf{a}_N(\psi)=\frac{1}{\sqrt{N}}\left[ 1,e^{j2\pi \frac{d}{\lambda} \psi},\dots,e^{j2\pi \frac{d}{\lambda} (N-1)\psi} \right],
	\label{ULA_steering_vector}
	\end{equation}
	where $d$ is the fixed antenna spacing with $d=\lambda/2$ and $\lambda$ denotes the carrier wavelength.
	
	\begin{figure}[!t]
		\centering
		\includegraphics[width=1\linewidth]{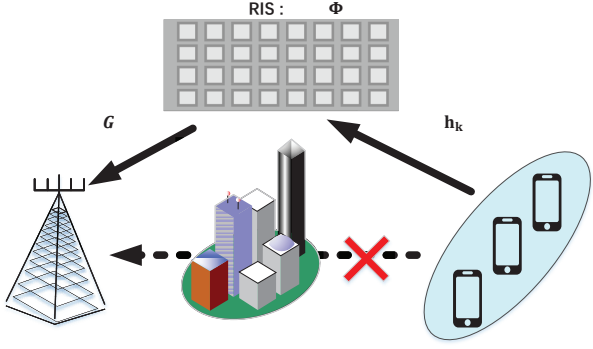}
		\caption{block diagram of RIS-aided MU-MIMO system.}
		\label{system model}
	\end{figure}

	In practical scenarios the total channel between UE and BS can be divided into direct LoS path and RIS-aided channel. Since the LoS path can be directly estimated when RIS turns off all reflective elements,  we neglect the LoS part here and only consider the RIS-aided channel estimation. Thus the received signals at BS can be written as follows
	\begin{equation}
	\arraycolsep=1.0pt\def\arraystretch{1.8}
	\begin{array}{lll}
	\mathbf{y}_{k,t}&=&\mathbf{G}\cdot \text{diag}(\bm\phi_t)\cdot\mathbf{h}_k s_{k,t}+\mathbf{n}_{k,t}\\
	&=&\mathbf{G}\cdot \text{diag}(\mathbf{h}_k)\cdot\bm\phi_t s_{k,t}+\mathbf{n}_{k,t}
	\end{array}
	\label{simu_model}
	\end{equation}
	
	In (\ref{simu_model}), $\bm\phi_t\in\mathbb{C}^{N_\text{I}\times 1}$ denotes the phase shifting vector of RIS at $t$ timeslot and $\mathbf{n}_{k,t} \sim \mathcal{CN}(\mathbf{0},\sigma_N^2\mathbf{I}_{N_\text{BS}})$ represents the additive white Gaussian noise (AWGN) at the receiver. After $T$ timeslots, we can combine the $T$ equations from $t=1$ to $T$ and rewrite (\ref{simu_model}) as
	\begin{equation}
	\mathbf{\hat{Y}}_k=\mathbf{G}\cdot \text{diag}(\mathbf{h}_k)\cdot \bm\Phi + \mathbf{N}_k,
	\label{comb_model}
	\end{equation}
	where $\mathbf{\hat{Y}}_k, \mathbf{N}_k\in\mathbb{C}^{N_\text{BS}\times T}$ are the combinations of $\mathbf{y}_{k,t}$ and $\mathbf{n}_{k,t}$ along the timeline respectively. With the property of sparsity for Saleh-Valenzuela model, we reformulate (\ref{comb_model}) via beamspace channel representation $\mathbf{\hat{G}}=\mathbf{F}\mathbf{G}\mathbf{F}^H$ and $\mathbf{\hat{H}}_k=\mathbf{F}\text{diag}(\mathbf{h}_k)\mathbf{F}^H$ ($\mathbf{F}$ is the FFT matrix) and we can obtain:
	\begin{equation}
	\mathbf{F}\mathbf{\hat{Y}}_k=\mathbf{\hat{G}}\cdot \mathbf{\hat{H}}_k\cdot \mathbf{F}\bm\Phi + \mathbf{N}_k,
	\label{beamspace_model1}
	\end{equation}
	
	Notice that $\mathbf{\hat{G}}\mathbf{\hat{H}}_k$ is the beamspace cascaded channel marked as $\mathbf{H}_k^H$. Let $\mathbf{Y}_k=\mathbf{\hat{Y}}_k\mathbf{F}^H$ denote measurement signals and $\mathbf{A}=\mathbf{\Phi}^H\mathbf{F}^H$ denote sensing matrix. Then we can reformulate it as conventional compressive sensing model for estimation:
	\begin{equation}
	\mathbf{Y}_k=\mathbf{A}\cdot \mathbf{H}_k + \mathbf{N}_k^H.
	\label{beamspace_model2}
	\end{equation}
	
	To estimate the sparse cascaded beamspace channel $\mathbf{H}_k$ with lower pilot overhead, CS theory can be utilized here such as the conventional single measurement vector (SMV) and multiple measurement vectors (MMV) problems \cite{MMV}. Unfortunately,  direct applications of those CS methods usually lead to performance loss due to the neglected useful prior information, i.e., the specific triple-structured sparsity. Consequently, more pilot overhead is required to guarantee enough estimation performance in conventional CS-based methods.
	
	\section{Structured Sparsity Analysis}
	
	\begin{figure*}[!t]
	\centering
	\includegraphics[width=0.9\linewidth]{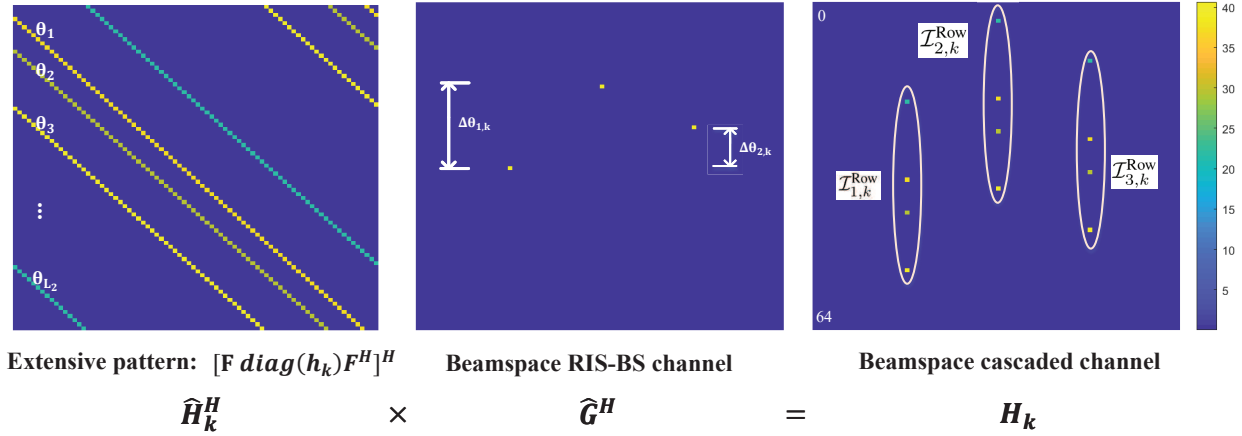}
	\caption{The beamspace cascaded channel $\mathbf{H}_k=\mathbf{\hat{H}}_k^H\times \mathbf{\hat{G}}^H$ and structured sparsity analysis.}
	\label{H_beam}
	\end{figure*}

	In this section, we illuminate the specific triple-structured sparsity in the cascaded beamspace channel $\mathbf{H}_k$. We depict the procedure for $\mathbf{H}_k=\mathbf{\hat{H}}_k^H\mathbf{\hat{G}}^H$ in Fig.2, where the darkness of color indicates the corresponding channel gains in the angular domain.
	
	As shown in Fig.2, the RIS-BS beamspace channel $\mathbf{\hat{G}}$ has both row and column sparsity and the UE-RIS beamspace channel $\mathbf{\hat{H}}_k=\mathbf{F}\text{diag}(\mathbf{h}_k)\mathbf{F}^H$ displays a generalized diagonal sparsity. It is uncomplicated to get that, assume only one path exists in $\mathbf{h}_k$ with AoA $\theta=0$ and path gain $\alpha=\alpha_0$, then the beamspace UE-RIS channel $\mathbf{\hat{H}}_k$ will be diagonal with diagonal elements $\mathbf{\hat{H}}_k(i,i)=\alpha_0$. When the AoA $\theta\neq 0$, it turns to a circulant-shift permutation matrix with shift distance as $N_\text{I}\sin(\theta)$. \footnote{We only consider on-grid situation and assume that $\sin{\theta}=n/N_\text{I},n\in\mathbb{Z}$ here. Off-grid estimation is left for works in future.} Generally, when there exist several paths in $\mathbf{h}_k$, the cascaded channel $\mathbf{\hat{H}}_k$ is composed of several circulant-shift permutation matrices, where each generalized diagonal corresponds to one uniform-quantized AoA as shown in Fig.2. And the triple-structured sparsity is analyzed in detail as follows:
	
	\begin{figure}[!t]
	\centering
	\includegraphics[width=0.9\linewidth]{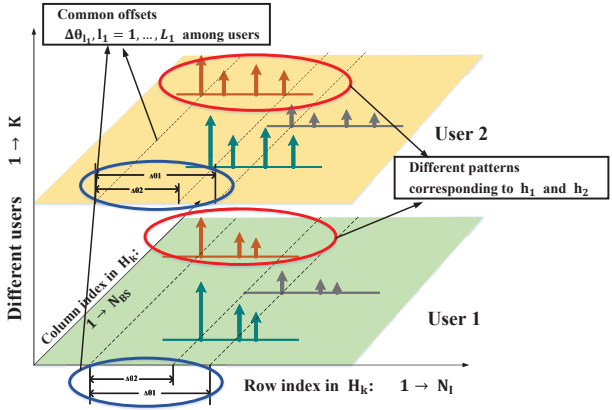}
	\caption{Block diagram of the triple-structured sparsity among users.}
	\label{ULA_sparsity}
	\end{figure}

\subsubsection{Column sparsity}

		\
		
		 Several previous studies have explored this sparsity such as \cite{chen2019channel,weixiuhong}, which is caused from the common sparse scatters at BS side. As shown in Fig.2, the beamspace cascaded channel $\mathbf{H}_k$ is column-sparse due to the limited AoAs $\theta_{l_1}^{\text{G}_r}$ in RIS-BS channel. Let $\mathcal{I}_k^{\text{Col}}$ denote the indices set of non-zero columns in $\mathbf{H}_k$, from \cite{chen2019channel} we can get that 
		\begin{equation}
		\mathcal{I}_{1}^{\text{Col}}=\mathcal{I}_{2}^{\text{Col}}=\dots =\mathcal{I}_{K}^{\text{Col}}, \ \ |\mathcal{I}_{k}^{\text{Col}}|=L_1.
		\label{colum_sparsity}
		\end{equation}
		
		This sparsity has been widely adopted before and we omit detailed illumination here for brevity.
		
		
\subsubsection{Structured row sparsity (after offset compensation)}

\
		
		The cascaded sparse channel $\mathbf{H}_k$ can be regarded as a diffusion of $\mathbf{\hat{G}}^H$, where each element is extended by scatters in UE-RIS channel. 
		We mark $\theta_{l_2,k}$ as the extensive pattern and regard the path locations of $\mathbf{\hat{G}}$ as initial points. For example in Fig.2, the three non-zero elements in $\mathbf{\hat{G}}^H$ are extended to three columns in $\mathbf{H}_k$, where the extensive pattern is directly controlled by the distribution of $\theta_{l_2},l_2=1,2,3,4$. Therefore we can get that, after compensating the offsets caused by different AoDs $\theta_{l_1}^{G_t}$, there exists a specific structured sparsity on the rows of cascaded beamspace channel $\mathbf{H}_k$. As marked by blue circles in Fig.3, the offset value for the $k$-th user in the $l_1$-th non-zero column is marked as $\Delta\theta_{l_1,k}$ compared with the first non-zero column.  Let $\mathcal{I}_{l_1,k}^{\text{Row}}$ denote the indices of non-zero values in the $l_1$-th non-zero column of $\mathbf{H}_k$. For example in Fig.2 ($N_\text{I}=64$), we have $\mathcal{I}_{1,k}^{\text{Row}}=\lbrace 20,35,38,50 \rbrace$, $\mathcal{I}_{2,k}^{\text{Row}}=\lbrace 4,19,22,34 \rbrace$ and $\mathcal{I}_{3,k}^{\text{Row}}=\lbrace 10,25,28,40 \rbrace$, and the offset values for each non-zero column are $\Delta\theta_{1,k}=0$, $\Delta\theta_{2,k}=-16$ and $\Delta\theta_{3,k}=-10$. Therefore the bias-structured sparsity with several offsets inside the $k$-th user can be formulated as
		\begin{equation}
		\mathcal{I}_{1,k}^{\text{Row}}-\Delta\theta_{1,k}=\mathcal{I}_{2,k}^{\text{Row}}-\Delta\theta_{2,k}=\dots=\mathcal{I}_{L_1,k}^{\text{Row}}-\Delta\theta_{L_1,k}
		\label{offset_sparsity}
		\end{equation}
		
		Besides, we can also analyze this type of sparsity via (\ref{RIS_BS_channel}) and (\ref{UE_RIS_channel}) mathematically. The beamspace cascaded channel with ULA configuration can be formulated as follows according to \cite{chen2019channel}:
		\begin{equation}
		\arraycolsep=1.0pt\def\arraystretch{1.8}
		\begin{array}{lll}
		\mathbf{G}\cdot \text{diag}(\mathbf{h}_k)&=&\displaystyle \sum_{l_1=1}^{L_\text{G}}\sum_{l_2=1}^{L_k}\alpha_{l_1}^\text{G}\alpha_{l_{2,k}} \mathbf{a}_{N_\text{R}}\left(\sin(\theta_{l_1}^{\text{G}_r})\right)\\
		&&\times \mathbf{a}_{N_\text{I}}\left(\sin(\theta_{l_1}^{\text{G}_t})-\sin(\theta_{l_2,k})\right)^H,
		\end{array}
		\label{real_model}
		\end{equation}
		And we can obtain similar conclusion via (10).
		

\subsubsection{Common offsets shared by all users}

\
		
		As analyzed above, bias-structured row sparsity exists inside the cascaded channel $\mathbf{H}_k$ for each user $k$. Moreover, from all users' joint point of view, as shown in Fig.3, it can be observed that all users share the common offsets corresponding to the sparse column index $l_1$. This is because the offset $\Delta\theta_{k,l_1}$ between the $1$-th and $l_1$-th columns is only related to AoDs at RIS side, and UE-RIS channel $\mathbf{h}_k$ has no influence on it. This structured sparsity here can be formulated mathematically as
		\begin{equation}
		\Delta\theta_{l_1,1}=\Delta\theta_{l_1,2}=\dots=\Delta\theta_{l_1,K}=\sin(\theta_{l_1}^{\text{G}_t})-\sin(\theta_{1}^{\text{G}_t}).
		\label{common_offset}
		\end{equation} 
		And (\ref{common_offset}) can be regarded as a beneficial prior information for multi-user cascaded channel estimation, especially when Structure 2 is exploited with unknown offsets to be estimated.

	\begin{remark}
		It is worth noting that the three types of structures are first jointly analyzed and exploited in this paper. Actually Structure 1 has been widely applied in \cite{chen2019channel,weixiuhong} and brings a high improvement in estimation accuracy. But unfortunately, in \cite{weixiuhong} it is impractical to assume that all users share certain common paths among UE-RIS channels $\mathbf{h}_k,k=1,\dots,K$. And in \cite{chen2019channel}, the structured sparsity is summarized as a common subspace projection into beamspace RIS-BS channel $\mathbf{\hat{G}}$, where the authors didn’t give a mathematically clear analysis. Actually the common subspace property, regarded as an intuitive description for Structure 3, is insufficient and can be further improved. Until now little research has exploited Structure 1, 2 and 3 jointly for enhanced estimation to our best knowledge. 
	\end{remark}

	\section{On-grid Cascaded Channel Estimation}
	In this section, based on the triple-structured sparsity analyzed in Section III, we propose a novel multi-user joint cascaded channel estimation scheme named as MTSCS-CE. Three-phase procedure is designed to apply Structure 1, 3 and 2 successively. The calculational complexity is close to conventional OMP method but pilot overhead can be remarkably reduced under the same NMSE accuracy. Our proposed algorithm in this paper focus on on-grid scenarios, while super-resolution estimation based on the triple-structured sparsity and the tradeoff between estimation accuracy and complexity will be further studied to overcome the energy leakage problem. 
	
\subsection{Sparse-column estimation: Phase 1}
	
	Since the common column sparsity (Structure 1) has been well studied before, we directly utilize methods in \cite{weixiuhong} for the common column support estimation as Phase 1. Calculate the total power of each column in measurement signals $\mathbf{Y}_k$. Since users' channels are all column-supported by the same AoAs at BS, we sum up powers along different users to jointly estimate the maximum $L_1$ supporting beams, which is written as follows:
	\begin{equation}
	\bm\Omega^c=\bm\Gamma\left(\text{diag}(\  \sum_{k=1}^K \mathbf{Y}_k^H \mathbf{Y}_k\ ),\ L_1 \right),
	\end{equation}
	The operator $\bm\Gamma(\bm v,L_1)$ represents to extract the column indices corresponding to the largest $L_1$ values in vector $\bm v$. When we obtain the column support set $\bm\Omega^c$, we neglect the rest $N_\text{I}-L_1$ columns and cut off the matrix column dimension to $L_1$, i.e., set $\mathbf{Y}_k\leftarrow \mathbf{Y}_k(:,\bm\Omega^c),\  \mathbf{H}_k\leftarrow \mathbf{H}_k(:,\bm\Omega^c)$ for the following row-structured estimation. 
	
\subsection{Offset-structured sparse-row estimation: Phase 2 }
	
	In this phase we assume the offset $\Delta\theta_{l_1}$ is already known. By exploiting Structure 2, the cascaded channel estimation problem for user $k$ can be reformulated as:
	\begin{equation}
	\arraycolsep=1.0pt\def\arraystretch{1.8}
	\begin{array}{clc}
	\displaystyle \min_{\mathbf{H}_k}& &\displaystyle\text{supp}(\mathbf{H}_k(:,l_1))\\
	\text{s.t.}&& C_1:\|\mathbf{Y}_k - \mathbf{A}\mathbf{H}_k \|_F^2\leq \epsilon\\
	C_2:&& \lbrace \text{supp}(\mathbf{H}_k(:,l_1))-\Delta\theta_{l_1} \rbrace^{\circ}_{N_\text{I}}=\mathcal{S}_k, \forall \  l_1=1,\dots,L_1
	\end{array}
	\end{equation}
	where the $\lbrace\mathbf{a}\rbrace^{\circ}_N=\mathbf{a}-\lfloor \mathbf{a}/N \rfloor\times N$ denotes the period-shift operator moving all elements of $\mathbf{a}$ into the interval $[1,N]$. $C_1$ confirms the accuracy of channel estimation and $C_2$ denotes that  user $k$'s all columns $\mathbf{H}_k(:,l_1)$ share a common pattern $\mathcal{S}_k$ through corresponding offsets $\Delta\theta_{l_1}$ (Structure 2), where the pattern $\mathcal{S}_k$ is controlled by UE-RIS channel $\mathbf{h}_k$ and offsets $\Delta\theta_{l_1}$ are controlled by AoDs in RIS-BS channel $\mathbf{G}$. 
	
	Then the supporting vectors among different columns in $\mathbf{H}_k$ can be regrouped depending on $\Delta\theta_{l_1},l_1=1,\dots,L_1$. In another word, if the $j$-th element in column $1$ is non-zero, there must exist a corresponding non-zero value at the $\lbrace j-\Delta\theta_1+\Delta\theta_i \rbrace _{N_\text{I}}^{\circ}$ row of $i$-th column due to the offset-structured sparsity. Therefore, let $\lbrace p,p+\alpha_1-\alpha_2,\dots,p+\alpha_1-\alpha_{L_k} \rbrace_{N_\text{I}}^{\circ}$ denote the $p$-th grouped indices set and we only need to search for the dominant $L_2^{(k)}$ grouped sets from $p=1,\dots,N_\text{I}$ to reconstruct the offset-structured sparse beamspace channel, which is quite similar to simultaneous OMP method.
	
\subsection{Multi-user joint offset estimation: Phase 3}
	
	However, notice that the particular offset-structured sparsity cannot be exploited directly in Phase 2 because actually the offsets are unknown, which means we should first estimate the offset values before Phase 2, or estimate the pattern $\mathcal{S}_k$ and offset values $\alpha_i$ simultaneously and iteratively. 
	
	From previous work \cite{chen2019channel}, a coarse estimated beamspace channel can be easily obtained, which can be utilized here as an initial resolution and marked as $\mathbf{H}^{(0)}_k$. Although under conventional OMP scheme the gains may keep a large bias compared with the real channel, the offset is little affected by OMP accuracy. Since the overall columns in $\mathbf{H}_k$ are shifted from a common pattern $\mathcal{S}_k$ when user index $k$ is fixed, we can take a periodic cross-correlation between the first column and every other column in $\mathbf{H}^{(0)}_k$ to obtain the offset $\Delta\theta_{l_1}$. Besides, notice that offset values are controlled by AoDs of RIS-BS channel $\mathbf{G}$, which is shared by all user $k=1,\dots,K$ (Structure 3). Thus we can process an enhancement with all users considered together as:
	\begin{equation}
	\Delta\theta_{l_1}=\bm\Upsilon^1\left(\sum_{k=1}^K\left[\mathbf{H}_k(:,1)\circledast^1 \mathbf{H}_k(:,l_1)\right]\right)
	\end{equation}
	where $\bm\Upsilon^1(\cdot)$ denotes an operator to extract 1-dim index of the element with the largest amplitude and $\circledast^1$ represents 1-dim periodic cross-correlation operator between two vectors. Finally, the detailed specific algorithm can be summarized in Algorithm 1, which contains the triple-structured sparsity in cascaded channel.
	
	\begin{algorithm}[htb] 
		\normalem
		\caption{Multi-user joint Triple-Structured CS-based Channel Estimation (MTSCS-CE)} 
		\label{alg1} 
		\begin{algorithmic}[1] 
			\REQUIRE  measurement matrices $\mathbf{Y}_k$, sensing matrix $\mathbf{A}$, coarse estimation results $\mathbf{H}^{(0)}_k$ via conventional OMP, sparsity $L_1$ and $L_2^{(k)}, k=1,\dots,K$

			\ENSURE beamspace cascaded channel $\mathbf{H}_k,k=1,\dots,K$
			
			
			\emph{\% Phase 1: Sparse column estimation}
			
			\STATE select the dominant $L_1$ column indices $\bm\Omega^c$ depending on the signal power via (12). Then set $\mathbf{Y}_k\leftarrow \mathbf{Y}_k(:,\bm\Omega^c)$ and $\mathbf{H}_k^{(0)}\leftarrow \mathbf{H}_k^{(0)}(:,\bm\Omega^c)$
			
			\emph{\% Phase 3: Multi-user joint offset estimation}
			
			\STATE  calculate offsets $\Delta\theta_{l_1}$ from coarse estimation results $\mathbf{H}_k^{(0)}$ via multi-user circular correlation in (14).
			
			\emph{\% Phase 2: Structured row estimation}
			\STATE $\mathbf{r}_{k,l_1}\leftarrow \mathbf{Y}_k(:,l_1)$; $\mathbf{H}_k=\bm 0$
			
			\FOR{$k=1$ to $K$}
			\FOR{$l_2=l$ to $L_2^{(k)}$}
			\STATE $\displaystyle \hat{p}\leftarrow\argmax_p \sum_{l_1=1}^{L_1}\left\|\mathbf{A}^H(:,p+\Delta\theta_{l_1})\mathbf{r}_{k,l_1}\right\|^2$
			
			\STATE $\mathcal{P}_k\leftarrow \mathcal{P}_k \cup \{\hat{p}\}$
			
			\FOR{$l_1=1$ to $L_1$}
			\STATE $\bm\Xi_{k,l_1}\leftarrow\{\mathcal{P}_k+\Delta\theta_{l_1}\}^{\circ}_{N_\text{I}}$
			
			\STATE $\mathbf{r}_{k,l_1}\leftarrow[\mathbf{I}-\mathbf{A}(\bm\Xi_{k,l_1})\mathbf{A}^{\dag}(\bm\Xi_{k,l_1})]\cdot\mathbf{Y}_k(:,l_1)$
			\ENDFOR
			\ENDFOR
			\ENDFOR
			
			\emph{\% Output}
			
			\STATE $\mathbf{H}_k(:,\bm\Omega^c(l_1))=\mathbf{A}^{\dag}(\bm\Xi_{k,l_1})\cdot\mathbf{Y}_k(:,l_1),$
			
			 $ \ \ \ \ \ \ \ \ \ \ \ \ \ \ \ \ \forall\ \  k=1,\dots,K,\ \  l_1=1\dots,L_1$

		\end{algorithmic}
	\end{algorithm}

	\section{Extension to UPA Model}
	Instead of ULA, RIS is more widely configured with UPA elements, whose size is marked as $N_1\times N_2$. Correspondingly, the steering vector at RIS side is reformulated as:
	\begin{equation}
	\mathbf{\bar{a}}(\theta,\phi)=\mathbf{a}_{N_1}(\theta)\otimes\mathbf{a}_{N_2}(\phi),
	\end{equation}
	where $\theta$ and $\phi$ denote the azimuth and elevation angles in the path. The DFT codebook for UPA is rewritten as $\mathbf{\bar{F}}=\mathbf{F}_{N_1}\otimes \mathbf{F}_{N_2}$. Channel and system models here are similar to those in ULA scenario after replacing $\mathbf{a}_N$ by $\mathbf{\bar{a}}$.
	
	\begin{figure}[!t]
	\centering
	\includegraphics[width=1\linewidth]{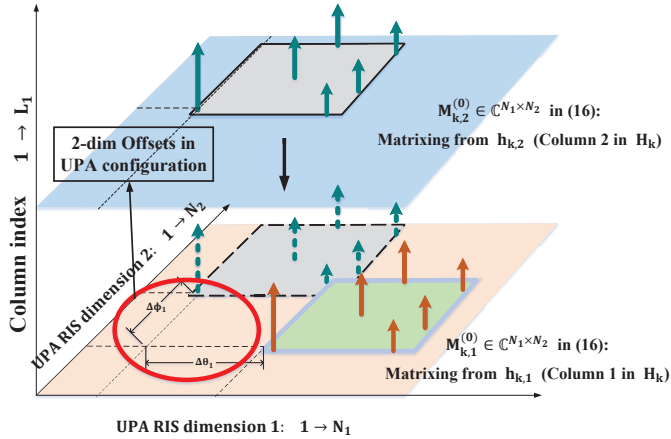}
	\caption{Block diagram of the triple-structured sparsity for single user under UPA configuration.}
	\label{UPA_sparsity}
	\end{figure}

	It is worth pointing out that the structured sparsity in UPA-RIS scenario is partially different from the analysis in Section III. The RIS-BS channel $\mathbf{G}$ can be similarly transformed into beamspace as a row-column double-sparse matrix. However, the beamspace sparsity for $\mathbf{\hat{H}}_k=\mathbf{\bar{F}}\text{diag}(\mathbf{h}_k)\mathbf{\bar{F}}^H$ changes. Due to the two azimuth and elevation periodic angles $\theta$ and $\phi$, the beamspace channel $\mathbf{\hat{H}}_k$ shows a two-hierarchical diagonal sparsity, which means that no evident column correlation exist in the final cascaded beamspace channel $\mathbf{H}_k$. 
	
	To handle this challenge and apply offset-structured sparsity property in UPA-RIS estimation, we rewrite the cascaded beamspace channel $\mathbf{H}_k\in\mathbb{C}^{N_1N_2\times N_\text{BS}}$ as a tensor form $\mathbf{H}'_k\in\mathbb{C}^{N_1\times N_2\times N_\text{BS}}$. As depicted in Fig.4, each column of $\mathbf{H}_k$ is reshaped to $N_1\times N_2$ and arranged horizontally. Then a direct 2-dim correlation clearly appears among different horizontal planes. 
	
	For on-grid triple-structured CS estimation, since the basic idea is similar to that in Section IV, we only demonstrate certain differences here. Assume the solution via OMP without structured property is known and marked as $\mathbf{H}_k^{(0)}=[\mathbf{h}_{k,1}^{(0)},\dots,\mathbf{h}_{k,N_\text{BS}}^{(0)}]$. First, we should take each non-zero column into one matrix $\mathbf{H}_{k,l_1}^{(0)}$, i.e. 
	\begin{equation}
	\mathbf{M}_{k,l_1}^{(0)}=\text{unvec}_{N_1,N_2}\left(\mathbf{h}_{k,l_1}^{(0)}\right),
	\end{equation}
	and then tune (14) from 1-dim correlation to 2-dim correlation:
	\begin{equation}
	(\Delta\theta_{l_1},\Delta\Phi_{l_1})=\bm\Upsilon^2\left(\sum_{k=1}^K\left[\mathbf{M}_{k,l_1}^{(0)}\circledast^2 \mathbf{M}_{k,1}^{(0)}\right]\right),
	\end{equation}
	where $\bm\Upsilon^2(\cdot)$ denotes an operator to extract 2-dim index of the element with the largest amplitude and $\circledast^2$ represents 2-dim periodic cross-correlation operator between two matrices. 
	
	When offsets are obtained via (17), we rearrange $N_1N_2$ supporting vectors in sensing matrix $\mathbf{A}$ to 2-dim $N_1\times N_2$ plane, marked as $\mathbf{v}_{n_1,n_2}$, and combine
	\begin{equation}
	\mathcal{S}_{n_1,n_2}=\{\mathbf{v}_{n_1+\Delta\theta_{1},n_2+\Delta\Phi_{1}},\dots,\mathbf{v}_{n_1+\Delta\theta_{L_1},n_2+\Delta\Phi_{L_1}}\}
	\end{equation}
	as a joint supporting set for the structured sparsity. Then sparse results can be achieved by greedily traversing all $N_1N_2$ set from $(n_1=1,n_2=1)$ to $(n_1=N_1,n_2=N_2)$ via simultaneous OMP.
	
	\section{Simulation Results}
	
	In this section we depict our simulation results for the proposed structured CS-based estimation scheme.
	Simulation results in the UPA scenario are similar to ULA and thus we only display ULA simulations here for brevity and space restriction.
	Assume $L_1=4$ paths exist in the RIS-BS channel $\mathbf{G}$ and path number in $\mathbf{h}_k$ is independently selected from $4$ to $8$, i.e., $4\leq L_2^{(k)}\leq 8$. Path gains follow complex Gaussian distribution, while only on-grid AoAs and AoDs are generated in the channels $\mathbf{G}$ and $\mathbf{h}_k$. In ULA configuration, we set $N_\text{BS}=64$, $N_\text{I}=128$ and $K=16$. 
	Signal-to-noise ratio (SNR) is set as $0\ $dB. Other detailed parameter settings can be found in \cite{weixiuhong}. Normalized mean square error (NMSE) is chosen as the accuracy metric for estimation.
	
	\begin{figure}[!t]
	\centering
	\includegraphics[width=1\linewidth]{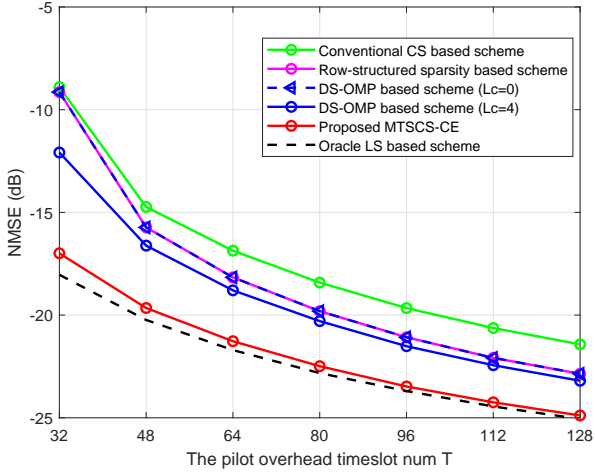}
	\caption{NMSE against pilot length $T$ with SNR fixed as $0$ dB.}
	\label{res_ULA}
	\end{figure}
	
	Fig.5 and Fig.6 compare the NMSE performance of the proposed MTPCS-CE algorithm, conventional CS based scheme \cite{OMP_conventional}, the row-structured
	sparsity based scheme \cite{chen2019channel}, DS-OMP method \cite{weixiuhong} and the oracle LS algorithm (lower bound). The row-structured-based scheme and DS-OMP method both adopt incomplete sparsity in the cascaded channel. The former only utilize the sparsity of BS AoAs and the latter is processed under a courageous assumption that all users share $L_c$ common paths in communication. In our simulation for convenience we set $L_c=4$. It is worth noting that this parameter $L_c$ only effect the performance of DS-OMP method but has no influence on other schemes including our proposed MTPCS-CE algorithm.
	
	Fig. 5 shows the impacts of pilot overhead timeslot number $T$ on the
	NMSE with SNR fixed to 0 dB. As shown in Fig.5, all estimation performances turn quite close to the oracle LS algorithm when pilot timeslots are large enough ($T\geq 128$), which can be illuminated via CS property with extremely large measurement. As for our proposed algorithm, it can be easily observed that the MTPCS-CE algorithm outperforms other partial sparsity-based schemes, since more helpful prior information (more sparse structures) are exploited in our proposed scheme. Consider $T=32$ and $\text{SNR}=0 \ \text{dB}$ as an example, we can find that the new algorithm  achieves an extreme enhancement about almost $10$ dB compared with DS-OMP scheme, and only keeps a quite close gap about $1$ dB to oracle LS method. The same conclusions can be obtained in the simulation of NMSE against SNR, as shown in Fig.6.

	\begin{figure}[!t]
	\centering
	\includegraphics[width=1\linewidth]{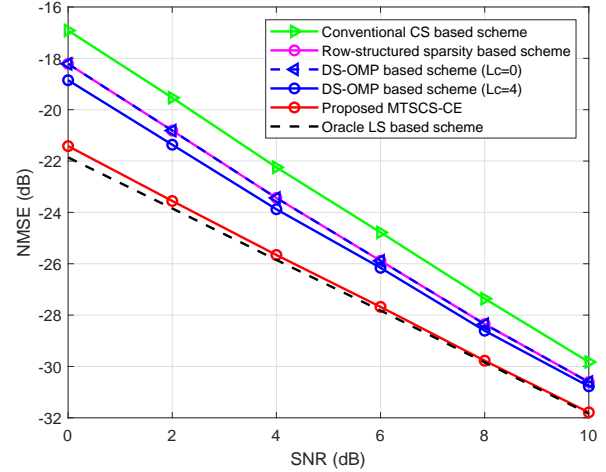}
	\caption{NMSE against SNR with pilot overhead length fixed as $T=64$ .}
	\label{res_UPA}
	\end{figure}
	
\section{Conclusion}
In this paper, by exploiting specific triple-structured sparsity of the cascaded channel, we propose a novel CS-based multi-user joint estimation algorithm. Sparse column supports and common offsets among users are estimated first. After offset compensation, the row-structured sparsity is further exploited via SOMP for estimation. Moreover, we extend our algorithm to UPA configuration, which is more practical in wireless communications. The complexity is quite low and simulation results show that the proposed approach can significantly reduce pilot overhead. 
	
	\section*{Acknowledgment}
	This work was supported in part by the National Key R$\&$D Program of China under Grant 2017YFE0112300.

	
	%

	%
	%
	%
	%
	%

	\ifCLASSOPTIONcaptionsoff
	\newpage
	\fi
	
	\bibliographystyle{ieeetr}
	\normalem
	\bibliography{bare_jrnl.bib}

	\vspace{12pt}

\end{document}